\documentclass[aps,prd,preprint,preprintnumbers,nofootinbib]{revtex4}

\usepackage{graphicx}
\usepackage[utf8]{inputenc} 
\usepackage{amsmath}
\usepackage{amssymb}
\usepackage{float}
\usepackage{comment}
\usepackage{soul}
\usepackage[usenames,dvipsnames]{color}

\definecolor{mightnightblue}{RGB}{25,25,112}

\definecolor{brown}{rgb}{0.59, 0.29, 0.0}

\renewcommand{\Re}{\operatorname{Re}} 
\renewcommand{\Im}{\operatorname{Im}}

\newcommand{\eps}{\varepsilon}
\newcommand{\tH}{\widetilde{H}}
\newcommand{\tU}{\tilde{U}}
\newcommand{\tlam}{\tilde{\lambda}}
\newcommand{\tJ}{\tilde{J}}
\newcommand{\tD}{\tilde{\Delta}}
\newcommand{\dm}[1]{\Delta m^2_{#1}}

\def\21{$\mathrm{SU(2)_L \otimes U(1)_Y}$}

\newcommand{\Cinvestav}{Departamento de F\'{\i}sica, Centro de
  Investigaci{\'o}n y de Estudios Avanzados del IPN\\ Apdo. Postal
  14-740 07000 Mexico, DF, Mexico}
  
\usepackage{hyperref}
 \hypersetup{
     colorlinks=true,
     linkcolor= Black,
     citecolor=mightnightblue,
     urlcolor=mightnightblue
     } 

\begin{document}

\title{Exploring NSI degeneracies in long-baseline experiments }
\author{L. J. Flores~$^1$}\email{jflores@fis.cinvestav.mx}
\author{E. A. Garc\'es~$^1$}\email{egarces@fis.cinvestav.mx} 
\author{O. G. Miranda~$^1$}\email{omr@fis.cinvestav.mx}

\affiliation{$^1$~\Cinvestav}

\begin{abstract}
 One of the main purposes of long-baseline neutrino experiments is to
 unambiguously measure the CP violating phase in the neutrino sector
 within the three neutrino oscillation picture. In the presence of
 physics beyond the Standard Model, the determination of the CP phase
 will be more difficult, due to the already known degeneracy
 problem. 
Working in the framework of non-standard interactions (NSI),
 we compute the appearance probabilities in an exact analytical
 formulation and analyze the region of parameters where this degeneracy 
 problem is present. We also discuss some cases where the degeneracy of the  
  NSI parameters can be probed in long-baseline experiments.
\end{abstract}

\maketitle
\section{Introduction}

Most of the Standard Model parameters in the leptonic sector have been
measured with high precision, including most of the mixing angles of
the PMNS
matrix~\cite{deSalas:2017kay,globalfit,Esteban:2016qun,Capozzi:2018ubv}
and the charged lepton masses~\cite{Patrignani:2016xqp}. It is
expected that
DUNE\cite{Acciarri:2015uup,Habig:2015rop,Acciarri:2016crz,Acciarri:2016ooe}
and Hyper-Kamiokande~\cite{Abe:2011ts,Abe:2015zbg} will accurately
measure the CP violating phase, $\delta$, if we restrict to the
standard three neutrino oscillation picture. The measurement of
absolute neutrino masses is another challenge, pursued by the Katrin
experiment~\cite{Angrik:2005ep}.

\noindent
On the other hand, the non-zero neutrino masses have motivated their
theoretical explanation through beyond the Standard Model physics. One
of the best motivated schemes is that of the
seesaw~\cite{Schechter:1980gr,Mohapatra:1980yp,GellMann:1980vs,minkowski:1977sc},
although there are plenty of beyond the Standard Model theories
searching to explain the neutrino mass pattern~\cite{Valle:2015pba}.
The presence of new physics leads naturally to a degeneracy on the
neutrino CP phases; for instance, non-unitarity of the leptonic mixing
matrix~\cite{Escrihuela:2015wra,Escrihuela:2016ube,Fong:2017gke,Tang:2017khg}
will lead to an ambiguity in the measurement of the standard CP
violating phase, $\delta$, as has been already pointed out
in~\cite{Miranda:2016wdr}. Models beyond the Standard Model also
include the sterile neutrino hypothesis, that has also been studied in
the context of long-baseline neutrino
experiments~\cite{Chatterjee:2017xkb,Dutta:2016glq,Choubey:2017ppj,Choubey:2017cba,Blennow:2016jkn}.

\noindent
A model independent framework aiming to incorporate a wide set of
models is the so called non-standard interaction (NSI)
picture~\cite{Farzan:2017xzy,Miranda:2015dra,Ohlsson:2012kf}, where
the information on new physics is encoded in parameters proportional
to the Fermi constant. Besides the search for new physics signals in
neutrino experiments, the robustness of the standard solution has also
been jeopardized by NSI~\cite{Miranda:2004nb} showing the importance
of short-baseline neutrino experiments that could help constrain these
parameters. Particularly, coherent elastic neutrino nucleus
scattering~\cite{Akimov:2017ade} has been helpful in obtaining this
restrictions~\cite{Kosmas:2017tsq,Esteban:2018ppq,Denton:2018xmq,AristizabalSierra:2018eqm}
as had been foreseen in~\cite{Barranco:2005yy}.

\noindent In this context, the sensitivity to NSI in the future DUNE
experiment~\cite{Acciarri:2015uup} has been extensively
studied~\cite{deGouvea:2015ndi,deGouvea:2016pom,Coloma:2015kiu,Coelho:2012bp,Deepthi:2016erc}
in order to know the expectative constraints in the future. It has
been found that, as in the non-unitary case, a degeneracy appears that
could weaken the resolution in the phase,
$\delta$~\cite{Forero:2016cmb}.  Due to this degeneracy, the
sensitivity of DUNE to the standard CP phase in presence of NSI has
been under
inquiry~\cite{Masud:2016bvp,Masud:2015xva,Ge:2016dlx,Liao:2016orc,Agarwalla:2016fkh,
  Das:2017fcz,Blennow:2016etl,Falkowski:2018dmy,Deepthi:2017gxg}.

\noindent
In this work we focus on the NSI framework in the context of
long-baseline neutrino experiments. We introduce an analysis of the
exact analytical formulas and will obtain useful information to search
for the regions leading to a degeneracy of the standard CP violating
phase, $\delta$, with the NSI parameters. We find the values of the
flavor-changing parameters that can mimic the standard appearance
probabilities, making the new phase, $\phi_{e\tau}$ indistinguishable
from $\delta$. We also discuss the implications of these values in the
biprobability plots, a very useful tool to exhibit the degeneracy
problem.  On the other hand, it is also interesting to find the
regions where a restriction to the NSI parameters can be done by
long-baseline neutrino experiments (LBNE).
It will be evinced that
biprobability plots can be used to search for these regions, although
in this case expectations are more limited.

\section{Nonstandard interactions in matter}
\label{sec:NSI}

New physics can affect the form of the different theories that
consider an extended gauge symmetry, additional number of fermion
singlets or extra scalars can be parametrized by the NSI
parameters~\cite{Ohlsson:2012kf,Miranda:2015dra,Farzan:2017xzy}.
Therefore, to study the effect of new physics in the neutrino matter
potential on Earth we will consider the NSI four-point effective
Lagrangian, whose coupling will be proportional to the Fermi
constant. In this way the non-standard interaction Lagrangian will be
given as
\begin{equation}
	\mathcal{L}_{NSI} =-2\sqrt{2}G_F\sum_f \eps^{f,P}_{\alpha\beta}\left[\bar{\nu}_\alpha
	\gamma^\rho L\nu_{\beta} \right]\left[\bar{f}\gamma_\rho Pf\right],
\end{equation}
where $f$ is a fermion of the first family ($e, u, d$) and $P$ is the
projector operator $P=L,R$.  In this work we will compute the effect
of charged leptons and neutrinos propagating in matter and, therefore,
we have taken $f=e$ and $\eps^u_{\alpha\beta}=\eps^d_{\alpha\beta}=0$.
To have an estimate of our results for the case of
$\eps^u_{\alpha\beta}$ (or $\eps^d_{\alpha\beta}$) one can consider
that the density of quarks on Earth is approximately three times that
for electrons~\cite{Escrihuela:2009up,Escrihuela:2011cf}.

\noindent
This new interaction has a non SM contribution to the neutrino-charged lepton
scattering process. As a consequence, neutrinos propagating in matter
will feel a new potential, additional to the usual charged-current
MSW~\cite{Wolfenstein:1977ue} potential. This can be introduced in the
propagation Hamiltonian and the total result will be 
\begin{equation} \label{eq:Hamiltonian}
	\tH = H + \frac{A}{2E}\left(\begin{array}{ccc}
		1+\eps_{ee} & \eps_{e\mu} & \eps_{e\tau}	\\
		\eps^*_{e\mu} & \eps_{\mu\mu} & \eps_{\mu\tau}	\\
		\eps^*_{e\tau} & \eps^*_{\mu\tau} & \epsilon_{\tau\tau}	
	\end{array} \right),
\end{equation}
where $H$ is the Hamiltonian in vacuum and the matter potential
$A=2\sqrt{2}G_FN_eE$, with $N_e$ is the electron number density and
$E$ is the neutrino energy.

\noindent
Due to the
NSI contribution, there are non-diagonal terms in the Hamiltonian. 
To study the impact of NSI interactions on long-baseline experiments,
wc compute here the exact expression for the oscillation probability
in matter. We briefly mention the already known standard case and
introduce the corresponding NSI formulas.  To make the expressions
more accessible to the reader, we show the flavor changing case for
$\eps_{e\tau}$ and set to zero all other NSI parameters.

\noindent
We will compute first the effective neutrino mass in matter. We will
follow the method used originally in~\cite{Zaglauer:1988gz} using an
approach that is independent of the
parametrization~\cite{Flores:2015mah}.  To find the exact expressions
for the effective squared masses $\bar{M}_i^2\equiv \tlam_{i}$ in the
presence of NSI, we start with the characteristic equation for the
Hamiltonian in Eq.~\eqref{eq:Hamiltonian}:

\begin{equation}
\label{eq:polynomial}
\tlam^3-\alpha \tlam^2 + \beta \tlam - \gamma=0,
\end{equation} 
\noindent whose real solutions are given by 
\begin{equation}
\label{eq:roots}
\tlam_n=\frac{\alpha}{3}+\frac{2}{3}\sqrt{\alpha^2-3\beta}\cos\left[\frac{1}{3}\arccos
\left( \frac{2\alpha^3-9\alpha\beta+27\gamma}{2\sqrt{(\alpha^2-3\beta)^3}}\right)+\frac{2(n-1)\pi}{3} \right], \quad n=1,2,3
\end{equation}

\noindent that in our case take the form
\begin{eqnarray} \label{2.12}
\alpha &=& \dm{21}+\dm{31}+A, \nonumber \\ 
\beta &=& \dm{31}\dm{21}+
A\dm{21}[1-|U_{e2}|^2 -2\Re(\eps_{e\tau}U^*_{e2}U_{\tau 2})] \nonumber \\ 
&+& A\dm{31}[1-|U_{e3}|^2 -2\Re(\eps_{e\tau}U^*_{e3}U_{\tau 3})] 
-A^2 |\eps_{e\tau}|^2, \\ 
\gamma &=& A\dm{21}\dm{31}[|U_{e1}|^2+2\Re(\eps_{e\tau}U^*_{e1}U_{\tau 1})]
\nonumber\\ 
&-&A^2|\eps_{e\tau}|^2(\dm{21}|U_{\mu 2}|^2 + \dm{31}|U_{\mu 3}|^2). \nonumber 
\end{eqnarray}
\noindent
In this equation, the NSI parameters introduce a new dependence on the
phases $\delta$ and $\varphi_{e\tau}$. This can be noticed, for
instance, by looking at the terms that go as $2\Re(\eps_{e\tau} U^*_{e
  i}U_{\tau i})$, that depend on the new phase, $\varphi_{e\tau}$.
The last quadratic term, $|\eps_{e\tau}|^2$, also introduces a new
dependence on $\cos\delta$ through $|U_{\mu 2}|^2$.

\noindent The previous relations in Eq. (\ref{eq:roots}) lead to three
eigenvalue equations corresponding to the effective squared masses
\begin{eqnarray}\label{eq:effectiveMasses}
\tlam_1 &=&  \frac{\alpha}{3}-\frac{1}{3}\sqrt{\alpha^2 -3\beta}\eta -\frac{\sqrt{3}}{3}\sqrt{\alpha^2 -3\beta}\sqrt{1-\eta^2}, \nonumber \\
\tlam_2 &=&  \frac{\alpha}{3}-\frac{1}{3}\sqrt{\alpha^2 -3\beta}\eta +\frac{\sqrt{3}}{3}\sqrt{\alpha^2 -3\beta}\sqrt{1-\eta^2},  \\
\tlam_3 &=&  \frac{\alpha}{3}+\frac{2}{3}\sqrt{\alpha^2 -3\beta}\eta ,  \nonumber 
\end{eqnarray}
where we have defined
\begin{equation}
	\eta = \cos\left[\frac{1}{3}\arccos\left( \frac{2\alpha^3-9\alpha\beta+27\gamma}{2\sqrt{(\alpha^2-3\beta)^3}}\right)\right] \nonumber .	
\end{equation}

\noindent
Once we have computed the effective masses in the NSI picture, we
proceed to compute the neutrino probabilities in terms of the mixing
matrix in this new basis, $\tU$. To make this computation, we
rearrange first the form of our Hamiltonian in
Eq. (\ref{eq:Hamiltonian}). This will make the appearance probability
expressions more readable.  Our main motivation is that, as it has
been shown, the biprobability plots have an elliptic shape when the
dependence of the oscillation probability on the CP violating phase
$\delta$ is considered~\cite{Kimura:2002hb}.  We will follow the same
procedure including now the dependence on the NSI parameters.

\noindent
For simplicity, in what follows we will show the
analysis for only one additional NSI parameter, the flavor changing
$\eps_{e\tau}$ and its phase $\varphi_{e\tau}$. 
Writing down the Hamiltonian from Eq.(\ref{eq:Hamiltonian}) as
\begin{equation}
	\tH = H + \frac{A}{2E}\mbox{diag}(1,0,0) + \frac{A}{2E}\eps,
\end{equation}
we can define two relations that will be useful later
\begin{eqnarray}\label{eq:definitionsH}
	\tH_{\mu e} &=& \frac{p}{2E}, \nonumber \\
	\tH_{\mu\tau}\tH_{\tau e} - \tH_{\mu e}\tH_{\tau\tau}
	&=&\frac{q}{(2E)^2} + \frac{A}{2E}r,
\end{eqnarray}
where 
\begin{eqnarray}
	\frac{p}{2E} &=& H_{\mu e}, \\
	\frac{q}{(2E)^2} &=& H_{\mu\tau}H_{\tau e} - H_{\mu e}H_{\tau\tau} ,\\
	r&=&H_{\mu\tau}\eps^*_{e\tau} .
\end{eqnarray}

\noindent
Note that these expressions have a similar form to the standard
case~\cite{Kimura:2002hb}, except for the additional dependence on $r$. 

\noindent
Both the vacuum Hamiltonian, $H$,
and the modified matter one, $\tH$, have the simple form
\begin{equation}
	H=\frac{1}{2E}UMU^\dagger, \quad \tH=\frac{1}{2E}\tU\tilde{M}\tU^\dagger,
\end{equation}
respectively, where $\tU$ is the modified matter mixing matrix, $M = \mbox{diag}(m^2_1, m^2_2, m^2_3)$, and $\tilde{M} = \mbox{diag}(\tlam_1, \tlam_2, \tlam_3)$. 
Taking these two expressions and  Eq.\eqref{eq:definitionsH}, one can find three relations for the product of $\tU_{\mu i}\tU_{ei}^*$:
\begin{eqnarray}
	\sum_i\tU_{\mu i} \tU^*_{e i}&=&\sum_i U_{\mu i}U^*_{ei}=0, \\
	\sum_i\tlam_{i}\tU_{\mu i} \tU^*_{e i}&=&\sum_i m^2_i U_{\mu i}U^*_{ei} = p, \\
	\sum_{(ijk)}^{cyclic}\tlam_{j}\tlam_{k}\tU_{\mu i} \tU^*_{e i}&=&\sum_{(ijk)}^{cyclic}m^2_j m^2_k U_{\mu i}U^*_{ei}
	+ 2EAr=q+2EAr.
\label{eq:recurrence}
\end{eqnarray}

\noindent
From here we can see that $p$ and $q$ are functions of the usual oscillation parameters. 
Solving this system of equations, we found the following relation
\begin{equation}\label{eq:productU}
	\tU_{\mu i} \tU^*_{e i}=\frac{\tlam_ip+q+2EAr}{\tD_{ji}\tD_{ki}}.
\end{equation}

\noindent
This product of entries of $\tU$ is important because it appears in
the oscillation amplitude for a muon neutrino to an electron neutrino:
\begin{equation}
	A(\nu_{\mu}\rightarrow\nu_e)=\sum_i \tU^*_{\mu i}\exp\left(-i\frac{\tlam_{i}L}{2E}\right)\tU_{ei}.
\end{equation}

\noindent
The oscillation probability for $\nu_{\mu}\rightarrow\nu_e$ is defined as the squared amplitude:
\begin{equation}
	P(\nu_\mu \to \nu_e)=|A(\nu_{\mu}\rightarrow\nu_e)|^2.
\end{equation}

\noindent
In terms of the Jarlskog invariant $J$ defined as $J=\Im(J^{12}_{\mu e})$, with $J^{ij}_{\alpha\beta}=U_{\alpha i}U^*_{\beta i}U^*_{\alpha j}U_{\beta j}$~\cite{Jarlskog:1985ht} 
and the  effective squared mass differences $\tD_{ij} = \tlam_{i}-\tlam_{j}=\bar{M}^2_i - \bar{M}^2_j$, we have
\begin{equation}\label{eq:appearanceProb}
	P(\nu_\mu \to \nu_e)=-4\sum_{(ij)}^{cyclic}
	\Re(\tJ_{\mu e}^{ij})\sin^2\left(\frac{\tD_{ij}L}{4E}\right) 
	-2\tJ\sum_{(ij)}^{cyclic}\sin\left(\frac{\tD_{ij}L}{2E}\right).
\end{equation}

\noindent
From equation~\eqref{eq:productU} we have
\begin{eqnarray}
\label{eq:Jarlskog}
	\Re(\tJ^{ij}_{\mu e})&=&\frac{\tlam_{i}\tlam_{j}|p|^2+|q+2EAr|^2
		+(\tlam_{i}+\tlam_{j})\Re[p(q^*+2EAr^*)]}{\tD_{ij}\tD_{12}\tD_{23}\tD_{31}} ,	\\
	\tJ &=& \frac{\Im[p(q^*+2EAr^*)]}{\tD_{12}\tD_{23}\tD_{31}}.
\label{eq:Jarlskog2}
\end{eqnarray}

\noindent
Let us notice that, if there were no NSI (meaning $r=0$), matter
effects would only appear in the effective masses
in Eq.~\eqref{eq:effectiveMasses} . 
The same would happen if we only have non-universal flavor-conserving NSI. 
On the other hand, the  non-diagonal NSI parameters have more 
complex effects due to their presence in Eq.~(\ref{eq:recurrence}).

\noindent
Replacing Eqs.~(\ref{eq:Jarlskog}) and
~(\ref{eq:Jarlskog2}) in 
Eq.~(\ref{eq:appearanceProb}), we can find the appearance probability
as a function of the CP phase $\delta$:

\begin{equation}
\label{eq:Papp}
P(\nu_\mu \to \nu_e) = a_1 +a_2 \cos\delta + a_3 \sin \delta + a_4 \cos 2\delta
 + a_5\sin 2\delta
\end{equation}
and similarly, for anti-neutrinos oscillation $\bar{\nu}_{\mu}\rightarrow \bar{\nu}_e$,
\begin{equation}
\label{eq:barPapp}
P(\bar{\nu}_\mu \to \bar{\nu}_e) = \bar{a}_1 + \bar{a}_2\cos\delta + \bar{a}_3 \sin \delta
+ \bar{a}_4\cos 2\delta + \bar{a}_5\sin 2\delta
\end{equation}
with the coefficients $a_k$ defined as
\begin{equation}
	a_k = \frac{-2}{\tD_{12}\tD_{23}\tD_{31}}
	\sum_{(ij)}^{cyclic}\left[\frac{2w^{ij}_k}{\tD_{ij}} 
	\sin^2\left(\frac{\tD_{ij}L}{4E}\right) + 
	y_k \sin \left(\frac{\tD_{ij}L}{2E} \right) \right],
\end{equation}
where $w^{ij}_{k}$ depends on all the standard oscillation parameters
and also on $\eps_{e\tau}$ and $\varphi_{e\tau}$, while $y_k$ is
independent from $\delta$.  The coefficients for the antineutrino
case, $\bar{a}_k$, have the same functional form, but with the changes
$A\rightarrow-A$, $\delta\rightarrow -\delta$, and
$\varphi_{e\tau}\rightarrow-\varphi_{e\tau}$.

\noindent
Equations~(\ref{eq:Papp}) and~(\ref{eq:barPapp}) have a similar 
form to the standard case. One difference is the appearance of the new
coefficients $a_4$ and $a_5$, although we have verified that
they are three orders of magnitude smaller than $a_1$, $a_2$, and
$a_3$. Another difference is that all coefficients now depend on
both phases. Again, we have verified that their variation is small, of
the order of few percent.

\noindent
With this exact formulation we will proceed, in the next chapter to
compute the relevant appearance probabilities to study the NSI picture
and to obtain the corresponding biprobility plots. As a cross check,
we have also computed our results using the approximate expressions
for  long-baseline NSI probabilities that have been considered
in~\cite{Barger:2001yr,Liao:2016hsa}. 
\section{NSI effects in long-baseline experiments}
After the previous description of the exact appearance probabilities
in the NSI framework, we can study the role of the NSI parameters
in long-baseline neutrino experiments, in particular for the
determination of the CP violating phase. 
Different works have studied the impact of NSI by comparing with the
appearance data~\cite{Coelho:2012bp,Friedland:2012tq} and have also
discussed the potential degeneracy with a new CP phase,
$\varphi_{e\tau}$, by analyzing either the expected survival
probability in the presence of NSI or the expected number of
events~\cite{Forero:2016cmb}.

\begin{figure}[h]
	\includegraphics[width=0.49\linewidth]{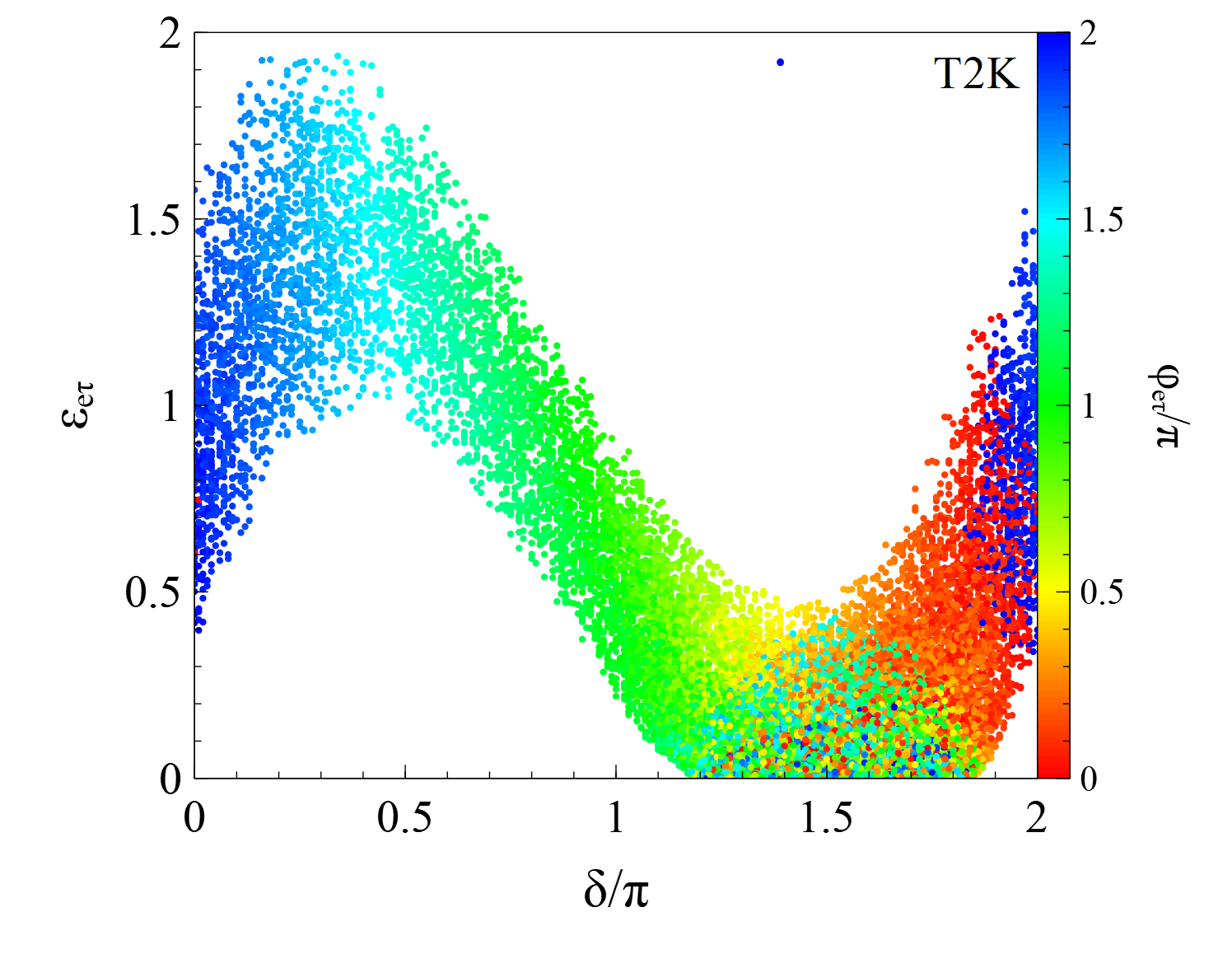} 
	\includegraphics[width=0.49\linewidth]{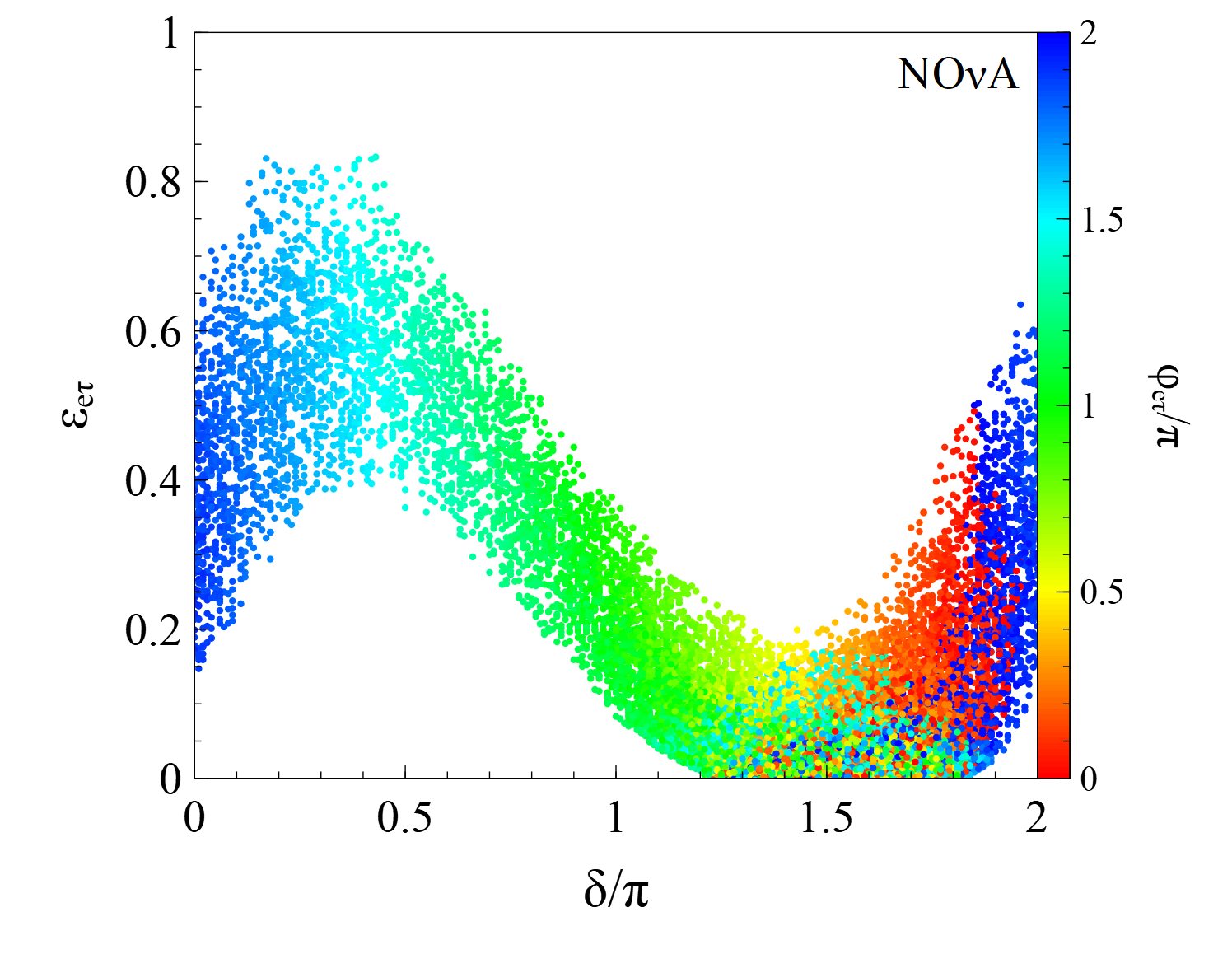} 
	\includegraphics[width=0.49\linewidth]{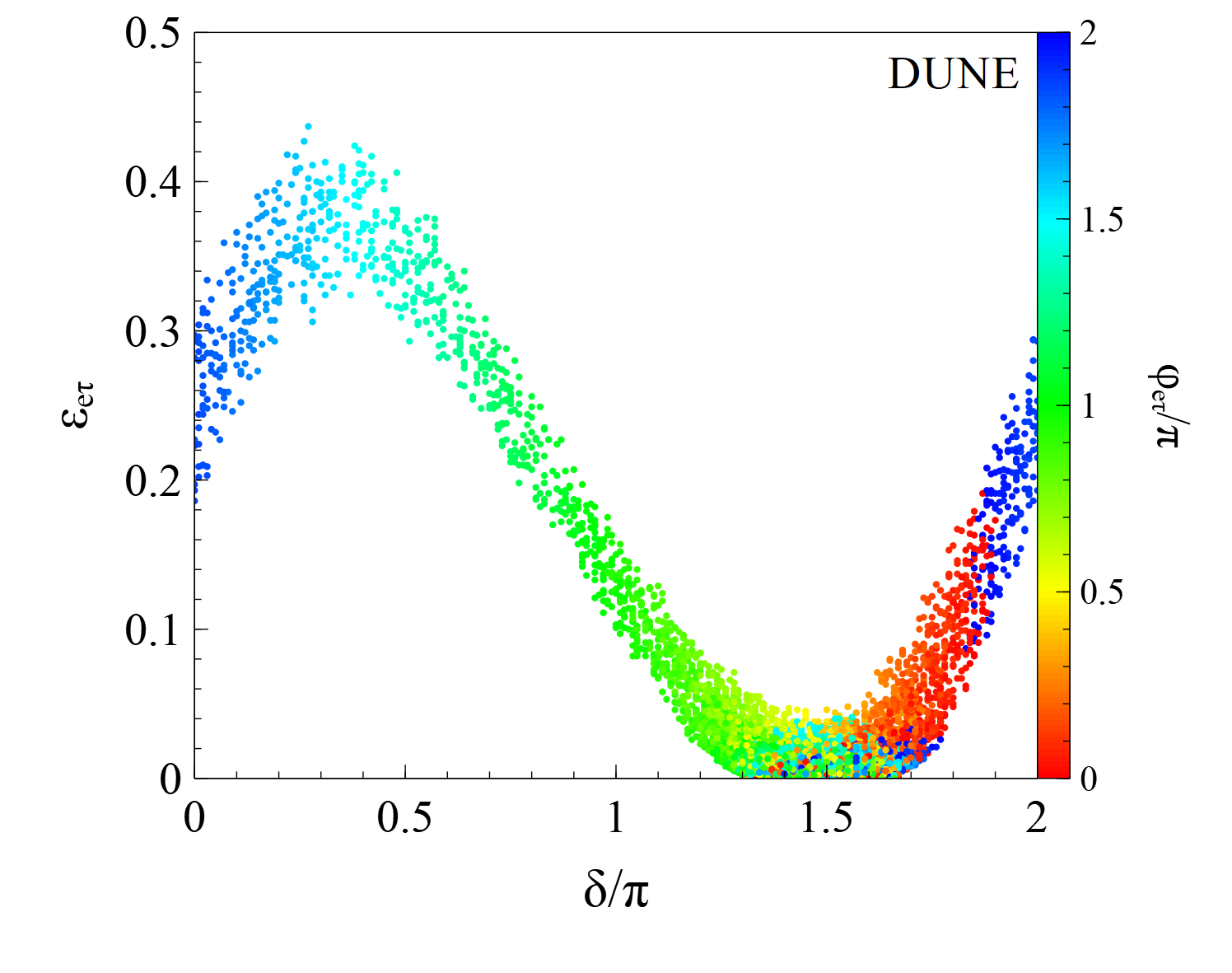} 
	\includegraphics[width=0.49\linewidth]{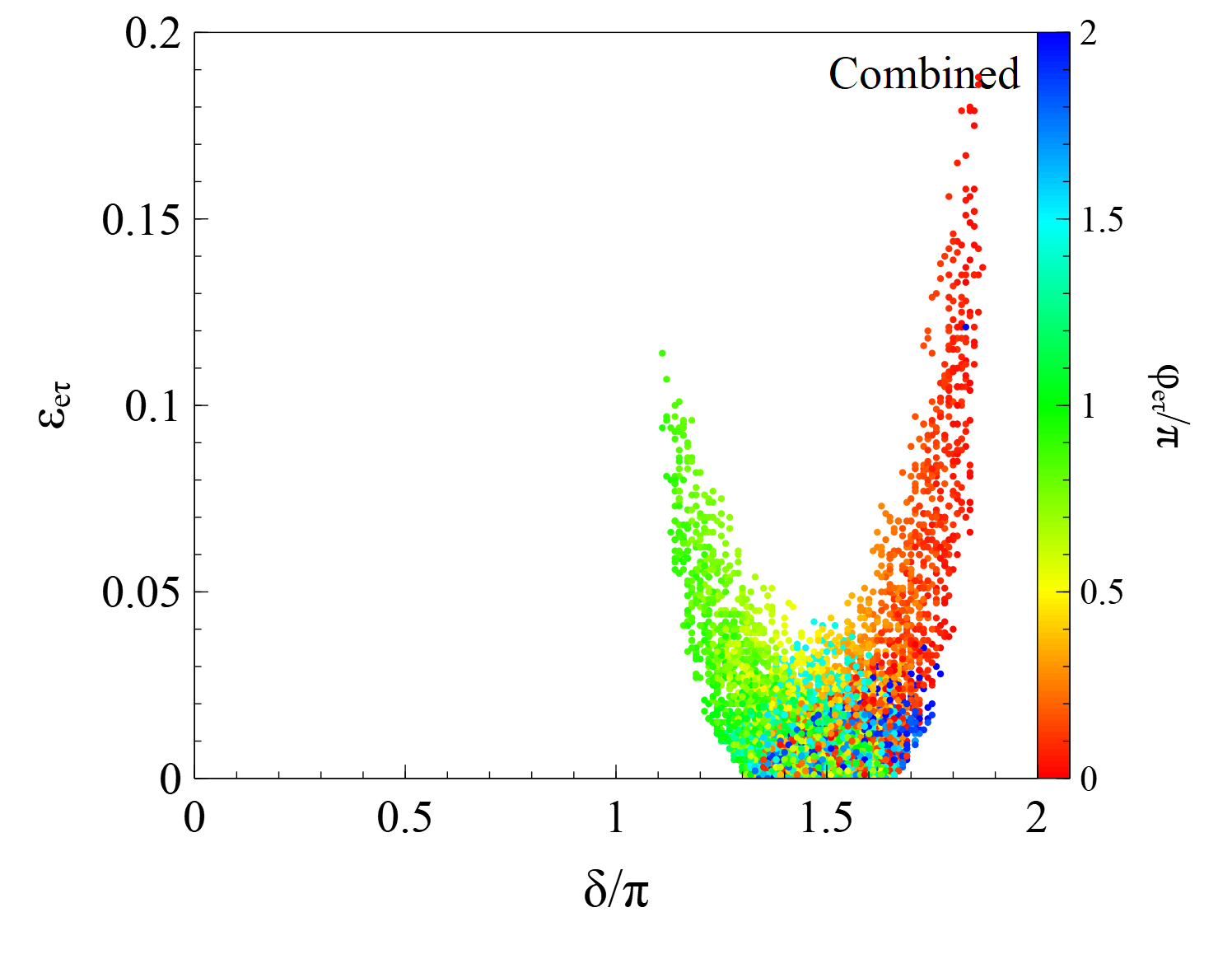} 
	\caption{Scatter plot of the standard CP violating phase,
          $\delta$, and the non-standard parameter $\eps_{e\tau}$ for
          a free non-standard CP phase, $\varphi_{e\tau}$. We show the
          dots that satisfy the biprobability region predicted by the
          standard oscillation picture. The first three panels show
          the case of T2K, NO$\nu$A and DUNE, while the bottom-right
          panel shows the values that satisfy simultaneously 
          the three experiments. The uncertainties that define 
          the appearance biprobability region, the baseline and the 
          average energy used for these plots are shown in 
          Table~\ref{table:biprobs}.}
	\label{fig:01}
\end{figure}

\noindent
We start our discussion by computing the NSI regions that would be
allowed by different long-baseline experiments. We consider the case
of the T2K collaboration~\cite{Abe:2017vif}, the NOVA
experiment~\cite{Adamson:2017gxd} and the future DUNE proposal that is
expected to measure the CP phase with high accuracy. This is shown in
Fig.~(\ref{fig:01}), where we also show the combined case for the
three experiments. We have made a scatter plot showing the points that
would be allowed for the three experiments. We have considered as a
test that the central value for the probabilities will be the one
corresponding to the standard case with a value of $\delta = 3\pi /2$
and we have assigned errors to the experiment's measurements according
to Table~\ref{table:biprobs}.  In the same table we have mentioned the
corresponding baseline and average energies considered for each
experiment.  In this scatter plot we have considered the central
values for the standard oscillation parameters~\cite{deSalas:2017kay},
a matter density of $\rho=2.84$ g/cm$^3$ and a constant electron
number density $N_e$. We show the different values of $\delta$,
$\eps_{e\tau}$, and $\varphi_{e\tau}$ that predict an allowed
probability for the corresponding case. We can see that for any
particular experiment there are different allowed points, leading to a
relatively small region when we consider the combination of the three
futuristic experimental results. Despite this, the 
  degeneracy region is still considerably large. 
It is important to mention that a more detailed analysis, considering
the neutrino spectrum for each experiment can reduce this degeneracy
region, especially for the futuristic case of DUNE, where a wide-band
beam neutrino flux will be used.

\begin{figure}[h]
	\includegraphics[width=0.51\linewidth]{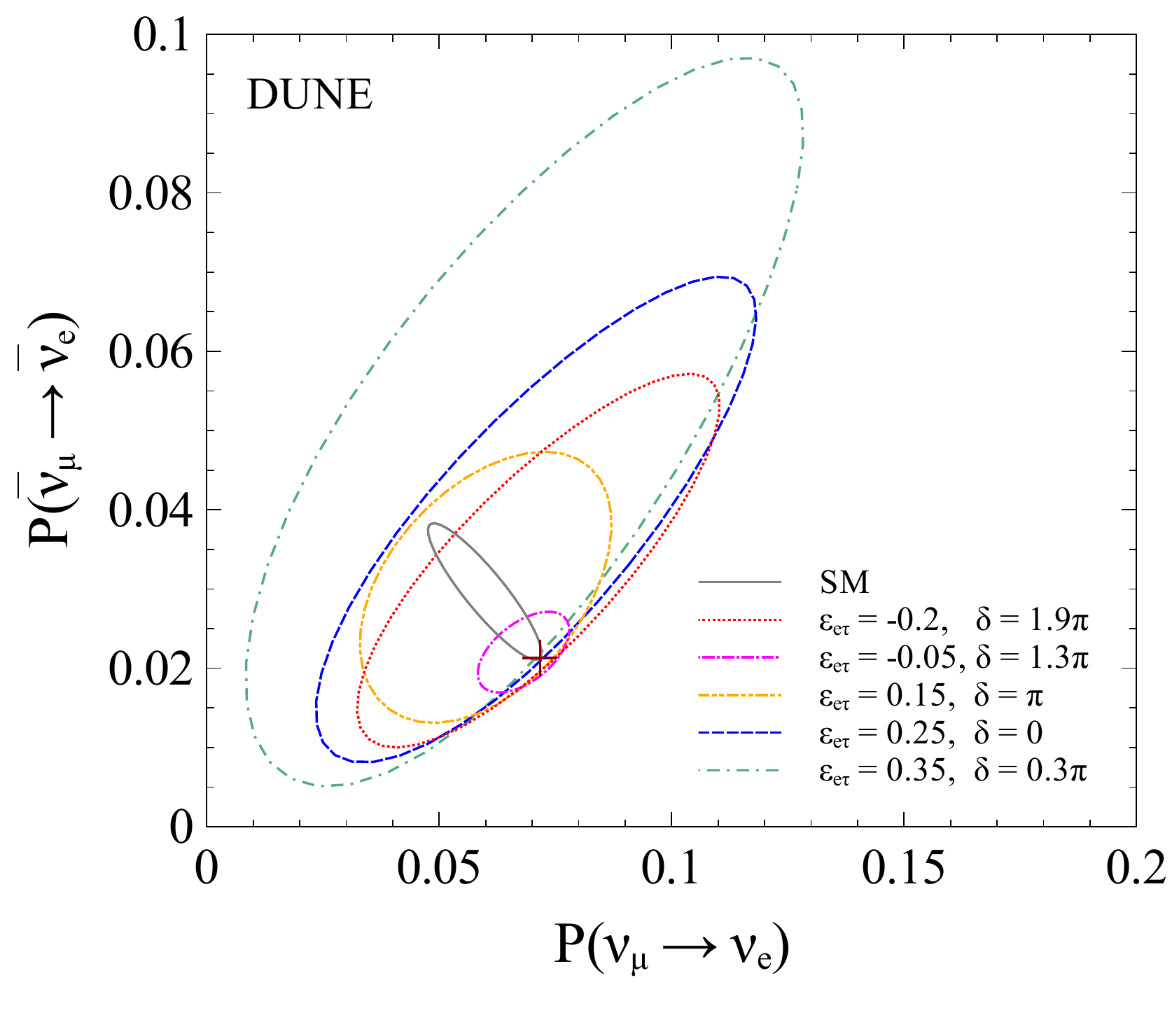} 
	\caption{Biprobability plots for the DUNE proposal, varying
          $\varphi_{e\tau}$ from $0$ to $2\pi$, for a fixed value of
          the NSI magnitude of $\eps_{e\tau}$ and the CP phase,
          $\delta$. The gray solid line represents the SM case, for
          varying $\delta$. Guided by the scatter plot from
          Fig.~(\ref{fig:01}), we find ellipses that pass through the
          appearance biprobability region (considering a value of
          $\delta=3\pi/2$) marked with a cross.  For the DUNE proposal
          we use $L=1300$ km and an average energy $E_\nu=3$~GeV. }
	\label{fig:02} 
\end{figure}

\noindent
The utility of this scatter plot, as a tool for the understanding of
the degeneracy regions, can be seen in Fig.~(\ref{fig:02}) where we
have considered the interesting case of the DUNE proposal as an
example.  As it is well known, biprobability plots can be studied to
have a general idea of the NSI parameters restrictions, or its
degeneracy.  In this figure we show the biprobabilities for fixed
values of $\delta$ and for the magnitude of the NSI parameter
$\eps_{e\tau}$.  These values were easily read from
Fig.~(\ref{fig:01}) and, as expected, the corresponding ellipses
always show a crossing point with the allowed region. The result is in
agreement with already reported cases~\cite{Forero:2016cmb} and can be
seen that many other values of $\eps_{e\tau}$ were easily found by
using the information from Fig.~(\ref{fig:01}).

\begin{table}[h]
	\centering
	\begin{tabular}{cccccccc}
		\hline\hline
		& \multicolumn{4}{c}{Uncertainties} & Baseline (km) & & Energy (GeV)     \\
		\hline\hline
		& $P(\nu_\mu \to \nu_e)$ & & $P(\bar{\nu}_\mu \to \bar{\nu}_e)$  & &  & &\\
		T2K   		& 10\%	& & 30\%      & &    295	& & 0.6  \\
		NO$\nu$A 	& 10\%	& & 25\%       & &   810	& & 2.0  \\
		DUNE  		& 5\%	& & 10\%       & &   1300	& &	3.0  \\
		\hline\hline
	\end{tabular}
	\caption{Expected uncertainties for the neutrino and
          anti-neutrino appearance probability for long-baseline
          neutrino experiment. In the last two column the characteristic
          baselines and average beam energy are shown. }
	\label{table:biprobs}
\end{table}

\begin{figure}[h!]
	\includegraphics[width=0.51\linewidth]{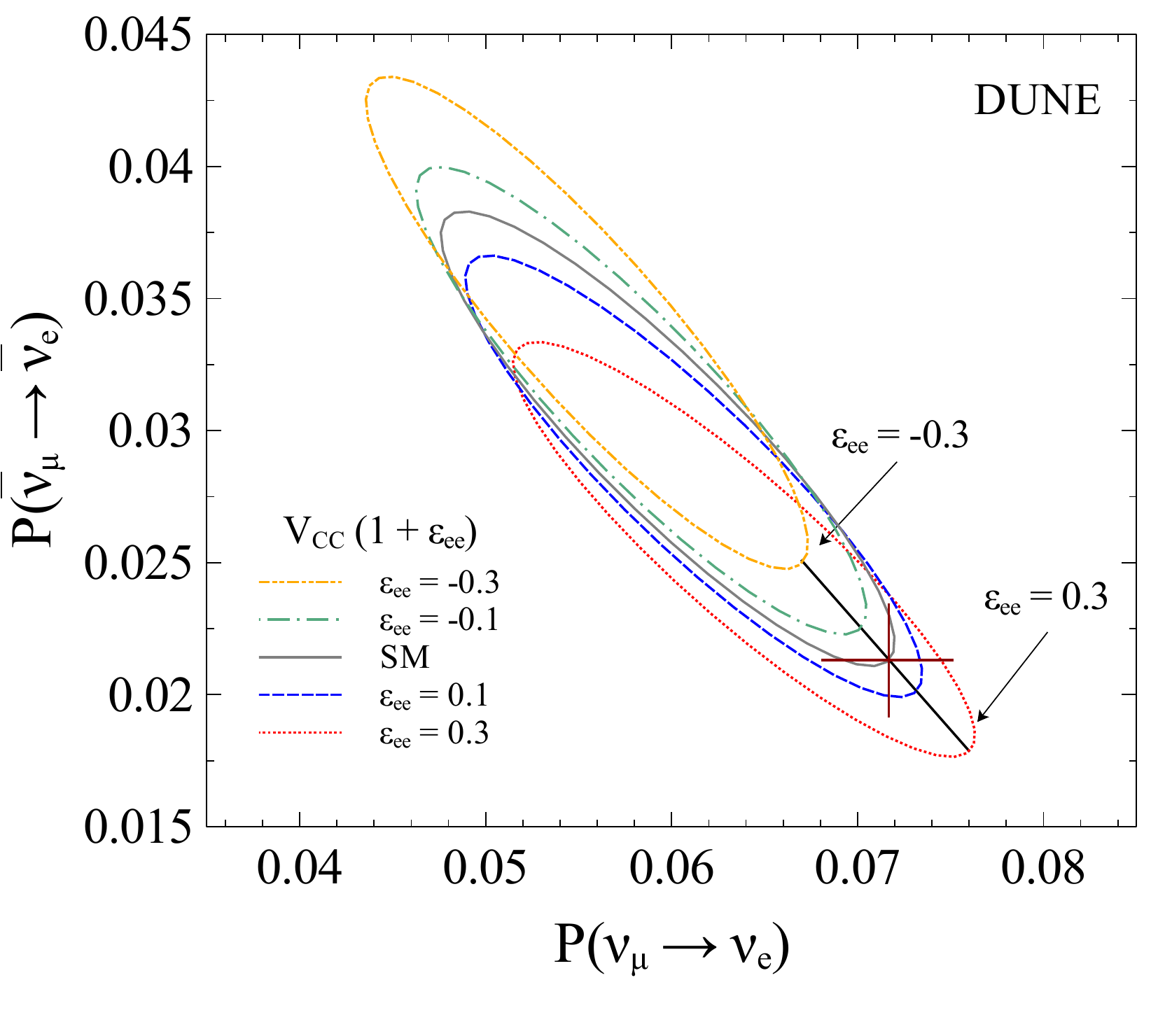}  
	\caption{Biprobability plots for the DUNE proposal,
          considering only $\eps_{ee}$ different from zero and varying
          $\delta$ from $0$ to $2\pi$. We show four ellipses, each for
          a different value of $\eps_{ee}$, along with the SM case
          (solid gray line). The appearance probability region for
          DUNE is marked with a cross, considering a value of
          $\delta=3\pi/2$. We consider again $L=1300$ km and an
          average energy $E_\nu=3$~GeV.  The presence of this diagonal
          NSI term only affects the standard charged-current potential
          as a small correction displacing the ellipse.}
	\label{fig:05}
\end{figure}

Another interesting analysis could be the search for restricted NSI
regions, instead of a degeneracy problem, in order to
look for future constraints from the DUNE experiment.
We separate this discussion in two natural cases, one involving the
presence flavor changing parameters, and the case of non-universal
terms. For the later case, we take $\eps_{ee}$ as the only parameter
different from zero. As a result, according to the discussion from
section~\ref{sec:NSI}, the NSI effects will be present only in the
effective masses. This implies an effective change in the potential:
$V_{CC} \to V_{CC}(1+\eps_{ee})$, resulting in a displaced ellipse of
the same size as in the standard oscillation picture when we vary
$\delta$. Several ellipses for this scenario are shown in
Fig.~\ref{fig:05}, along with a curve for fixed $\delta=3\pi/2$ and a
varying $\eps_{ee}$. Therefore, in this case a test of
the diagonal NSI parameter seems to be possible by long-baseline
neutrino experiments.

\begin{figure}[h]
        \includegraphics[width=0.49\linewidth]{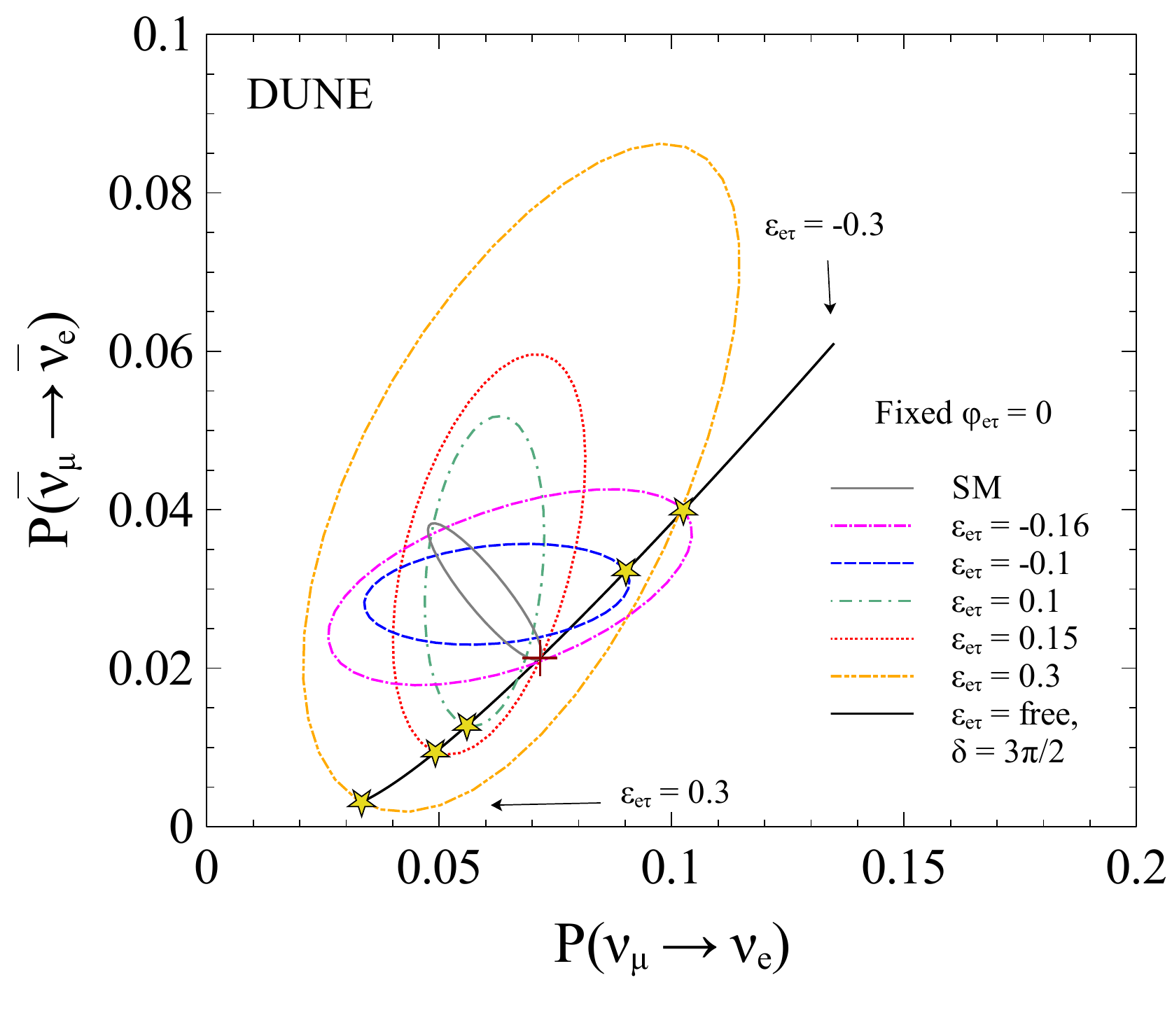} 
        \includegraphics[width=0.49\linewidth]{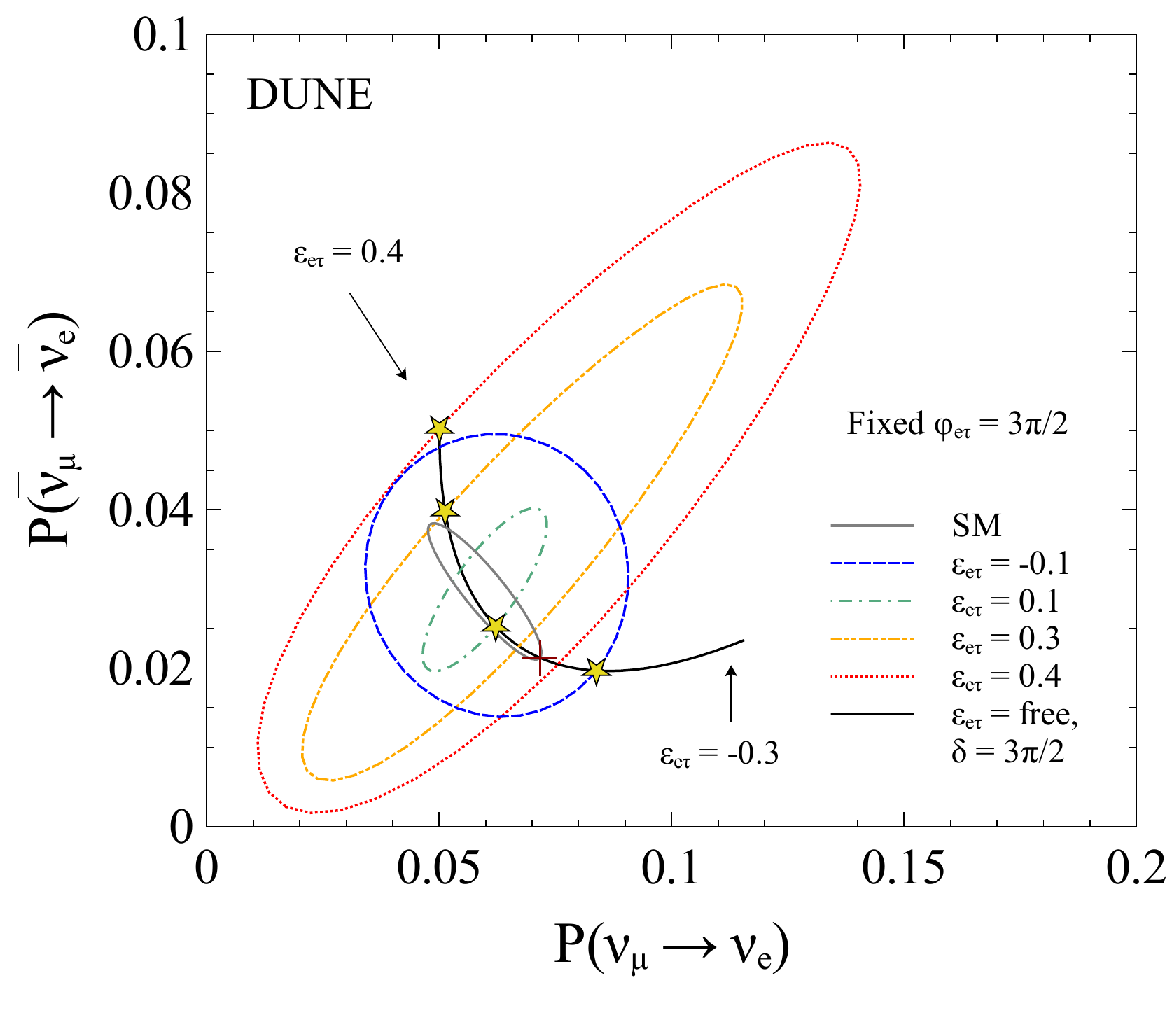} 
    \caption{Biprobability plots for the DUNE proposal, varying
      $\delta$ from $0$ to $2\pi$, for particular values of
      $\eps_{e\tau}$ and for $\varphi_{e\tau}=0\, (3\pi/2)$ for the
      left (right) panel.  The gray solid line stands for the SM case,
      and its prediction at $\delta=3\pi/2$ is shown with a cross,
      including its uncertainties, which are displayed in
        Table~\ref{table:biprobs}.  In both panels we can see that
      different values of $\eps_{e\tau}$ change the orientation and
      size of the ellipse.  The black line corresponds to a fixed
      value of $\delta=3\pi/2$ and $\varphi_{e\tau}=0\, (3\pi/2)$,
      while varying $\eps_{e\tau}$ in the range $[-0.3,0.3]$,
      ($[-0.3,0.4]$) . In this  black curve the yellow stars
      show its intersection with the ellipses.  }
    \label{fig:con} 
\end{figure}

On the other hand, for the case non-diagonal NSI parameters, we show
in Fig.~(\ref{fig:con}) the biprobability curves for $\eps_{e\tau}$
different from zero, varying the value of $\delta$. Since
$\eps_{e\tau}$ is a non-diagonal term, a new CP violating phase
$\varphi_{e\tau}$ might appear. For this reason, we present two cases:
$\varphi_{e\tau} = 0$ in the left panel and $\varphi_{e\tau} =3\pi/2$
in the right one. As explained in the previous section,
flavor-changing NSI modifies in a more complex way the oscillation probabilities
and, consequently the size and orientation of the biprobability
ellipses change notoriously, as seen in Fig.~(\ref{fig:con}).

We can notice here that the situation is more complicated than for the
diagonal NSI, making the restriction of the NSI parameters a more
complicated task. For instance, for the case of a $\varphi_{e\tau} =
0$, despite the particular value of $\delta=3\pi/ 2$ is shifted to a
region different from the Standard Model prediction, a different value
in the same ellipses can reach this region, allowing for a confusion
for a given value of $\eps_{e\tau}$. As expected, the quantitative
values of $\eps_{e\tau}$ and $\delta$ can be traced in the scatter
plot shown in Fig.~(\ref{fig:01}). For the case of $\varphi_{e\tau} =
3\pi /2$, it is possible to see that the perspectives for a NSI
restriction in this particular value are very promising as there are
almost no crossing points of the NSI ellipses with the biprobability
region, except for the particular case of a large NSI effect around
$\eps_{e\tau} = 0.4$.

\section{Conclusions}

In the standard three-neutrino oscillation picture, long-baseline
neutrino experiments will measure the mixing parameters with precision
and accuracy. In the presence of new physics the robustness of such
measurements is not guaranteed and different degeneracies may appear,
such as the well known LMA-D solution~\cite{Miranda:2004nb}.

\noindent
For the
determination of the CP phase, a similar problem has been pointed
out~\cite{Forero:2016cmb} when considering the flavor-changing NSI
parameter $\epsilon_{e\tau}$. In this case, again, the non-oscillatory
experiments will be of great help. In this work, we have focused in
the interplay of different long-baseline experiments. We have shown
the parameter space that will lead to an indetermination of the
$\delta$ value, as well as the role of a combined restriction from
several experiments. In all our computations, we have used an exact
formulation, discussing its main characteristics.

\noindent
The combination of different baselines can indeed help reduce the
degeneracy problem, although a more detailed study is needed. Besides,
we have computed the biprobability plots in the context of NSI and
prove its usefulness to understand the degeneracy problem in the
determination of the CP violating phase, when new physics is
present. We have illustrated this with the case of the future
experiment, DUNE. Although the combination of different baselines, and
the wide-band beam for the DUNE neutrino flux, could help in the
robust determination of the CP violating phase, short distance
non-oscillatory experiments seem necessary to better constrain the NSI
parameters.

\begin{acknowledgments}

This work was supported by CONACYT- Mexico and SNI 
(Sistema Nacional de Investigadores).

\end{acknowledgments}

\end{document}